\documentclass[12pt]{iopart}
\usepackage[dvips]{graphicx}
\usepackage{amssymb}
\usepackage{amsfonts}

\begin{document}

\title
[Demixing in colloid--polyelectrolyte star mixtures]
{Fluid-fluid demixing transitions\\ in colloid--polyelectrolyte star mixtures}

\author{Martin Konieczny\footnote{E-mail: 
kon@thphy.uni-duesseldorf.de} and Christos N Likos}

\address{
Institut f\"ur Theoretische Physik II: Weiche Materie\\
Heinrich-Heine-Universit\"at D\"usseldorf\\
Universit\"atsstra{\ss}e 1, 
D-40225 D\"usseldorf, Germany}

\begin{abstract}
We derive effective interaction potentials between hard, spherical 
colloidal particles and star-branched polyelectrolytes of various
functionalities $f$ and smaller size than the colloids. The 
effective interactions are based on a Derjaguin-like approximation,
which is based on previously derived potentials acting between
polyelectrolyte stars and planar walls. On the basis of these
interactions we subsequently calculate the demixing binodals 
of the binary colloid--polyelectrolyte star mixture, employing
standard tools from liquid-state theory. We find that the mixture
is indeed unstable at moderately high overall concentrations.
The system becomes more unstable with respect to demixing as 
the star functionality and the size ratio grow.
\end{abstract}

\pacs{61.20.-p, 61.20.Gy, 64.70.-p}


\section{Introduction}
\label{intro:sec}

Polyelectrolyte stars (PE's) are complex macromolecules that have
attracted a lot of interest in the recent past. They consist of 
$f$ polymer chains, all attached on a common centre, and carrying
ionizable groups along their backbones. Solution of these molecules
in a polar solvent results into dissociation of the groups, so that
the chains turn into polyelectrolytes and stretch considerably with
respect to their neutral conformations. Already in the early 1990s,
the importance of the stretched PE-chains in stabilizing colloidal
suspensions has been pointed out and analysed theoretically by
Pincus \cite{pincus:macromol:24} employing scaling theory
as well as, more recently, by Wang and Denton
\cite{wang:pre:04} using linear-response theory. A distinguishing feature of
PE-stars is their ability to adsorb the vast majority of the
released counterions into their interior, creating thereby an
inhomogeneous cloud of entropically trapped particles that
provides a strong entropic barrier against coagulation 
\cite{pincus:macromol:24,wang:pre:04,borisov:97,borisov:98,klein:99,klein:02,jusufi:prl:02,jusufi:jcp:02,denton:03}. The development of accurate effective
interactions between 
the PE-stars \cite{klein:02,jusufi:prl:02,jusufi:jcp:02,denton:03}
has led to predictions regarding
their overall phase behaviour with emphasis on crystallization
\cite{hoffmann:jpcm:03,hoffmann:jcp:04}, which has recently received
experimental corroboration \cite{ishizu:05}.

Though a great deal has thus been learned regarding the behaviour
of one-component solutions of PE-stars, the question of the influence
of these ultrasoft colloids on solutions of hard colloids has not
been investigated thus far. At the same time, the behaviour of 
PE-stars in the vicinity of planar or curved hard surfaces (such
as a larger colloidal particle) is an issue of considerable interest,
due to the possibility of manipulating the conformation of the 
PE-star by suitably changing the surface's geometry or physical
characteristics \cite{stamm:03, serpe:04,kim:05}. 
Recently, the properties of PE-stars
close to hard, planar walls were investigated in detail by means
of computer simulations and theory \cite{konieczny:jcp:06}. It has
been found that the geometrical constraint of the planar wall does
not affect the ability of the PE-stars to absorb the vast majority
of their counterions. In addition, a new mechanism giving rise to
a wall-star repulsion has been discovered, which rests on compression
of stiff star chains against the neighboring wall. In this work,
we proceed to the full, many body problem of a collection of 
PE-stars and neutral colloids, which can be seen as curved walls.
Basing on the results of Ref.\ \cite{konieczny:jcp:06}, we investigate
the structure of the mixture and find that it is unstable against
demixing as the concentration becomes sufficiently high. This work
serves, thereby, as the reference point for future investigations
on the effects of adding charge to the colloidal particles. It is
specular to recently published work on mixtures of {\it charged}
colloids with {\it uncharged} polymers \cite{tuinier:jpcm:05}, 
since in our case the colloids are neutral and the (star-branched)
polymers are charged.

The rest of this paper is organised as follows: in sec.\ \ref{sec:eff_pot}
we introduce the colloid--colloid and PE-star--PE-star effective
interactions and we derive the cross interaction, based on previous
results on the PE-star interaction potential with a planar wall.
In sec.\ \ref{sec:iet} we present our method for calculating
structure and thermodynamics by employing the aforementioned
interactions in combination with two-component liquid integral
equation theories. In sec.\ \ref{sec:res} we present our results
for various regimes of the parameter space as well as the overall
phase diagrams of the mixture. Finally, in sec.\ \ref{sec:summary}
we summarize and draw our conclusions.

\section{Effective pair potentials}
\label{sec:eff_pot}

The system under investigation is a binary colloid--PE-star mixture.
The colloids are coded with the subscript `c' and the PE-stars with `s'.
The mixture contains, thus,  
$N_{\rm c}$ spherical, neutral colloids with diameter $\sigma_{\rm c}$ 
(radius $R_{\rm c})$ and $N_{\rm s}$ PE-stars in aqueous solution. The stars can 
be characterised 
by their degree of polymerization $N_{\rm p}$, functionality $f$, and charging 
fraction $\alpha$. Thereby, 
the $f$ chains of each star are charged in a periodical 
manner in such a way that every ($1/\alpha$)-th monomer carries a charge $e$. 
As a result, every star carries a total bare charge 
$Q=e\alpha fN_{\rm p}$, leaving
behind $M=\alpha fN_{\rm p}$ monovalent, oppositely charged counterions in the 
mixture due to the requirement that the system must remain electro-neutral as a 
whole. With $\sigma_{\rm s}$ 
referring to the stars' diameter, i.e., 
twice the average centre-to-end distance $R_{\rm s}$ of the arms,
we define the size ratio $q$ between the two species as 
\begin{equation}
q=\sigma_{\rm s}/\sigma_{\rm c}.
\label{sizeratio:eq}
\end{equation}
Within this work, we will only consider PE-stars that are smaller than
the colloids, hence $q < 1$. The degree of polymerization of every arm,
$N_{\rm p}$, and the 
charging ratio $\alpha$ play a crucial role because 
they determine the number of 
released counterions $M$ mentioned above. The latter are, in turn, 
mainly responsible for the emergence of the 
star-star \cite{pincus:macromol:24,jusufi:prl:02,jusufi:jcp:02}
and the star-colloid
effective repulsions \cite{konieczny:jcp:06},
due to the loss of entropy they experience
when two such objects approach close to each other, see also 
eq.\ (\ref{vss:eq}) in what follows. In this work, we fix
$N_{\rm p} = 50$ and $\alpha = 1/3$ throughout. Generalizations
to other values of $\alpha$ and $N_{\rm p}$ can follow by 
appropriately taking into account the dependence of $M$ on these
parameters. Thereby, the two remaining single-molecule
parameters that we vary are
the stars' functionality $f$ and the size ratio $q$.

The thermodynamic parameters are 
the partial number densities 
$\rho_i=N_i/V$ ($i={\rm c},{\rm s}$)
of the 
respective species and the absolute temperature $T$. Alternatively, 
we can work with the concentrations 
$x_i=N_i/N$ and the total number density $\rho = N/V$,
with the total particle
number $N=N_{\rm c}+N_{\rm s}$   
in the overall volume $V$ of our model system.
We will consider constant,
room temperature ($T = 300\,{\rm K}$) throughout this work. 
This is the temperature for which the 
star-star effective interactions \cite{jusufi:prl:02,jusufi:jcp:02}
and the PE-star--planar wall potentials \cite{konieczny:jcp:06} have been 
derived, based on the value 
$\lambda_{\rm B} = 7.1\,{\rm \AA}$ for the Bjerrum length in
aqueous solvents. As usual, we define the
inverse thermal energy $\beta=1/(k_{\rm B}T)$,
with $k_{\rm B}$ denoting Boltzmann's
constant.

The starting point for all considerations to follow are the effective 
pair potentials between the constituent mesoscopic particles, having integrated
out all the monomer, solvent and counterions degrees of freedom. 
When introducing this set
of interactions as an input quantity into the full two-component integral 
equation theory described in more detail in Sec.\ \ref{sec:iet}, we 
can in principle completely access the 
structure and thermodynamics of the system at hand.

\subsection{The colloid--colloid and PE-star--PE-star interactions}
\label{sec:eff_pot1}

The effective colloid--colloid interaction at centre-to-centre distance $r$
is simply taken to be a pure hard sphere (HS) potential, namely:
\begin{equation}
\beta V_{\rm cc}(r)=
\left\{
\begin{array}{ll}
\infty & r\leq\sigma_{\rm c};\\
0 & {\rm else}.
\end{array}
\right.
\end{equation}

A lot of work concerning effective PE-star--PE-star interactions was done 
in the recent past by Jusufi and co-workers \cite{jusufi:prl:02,jusufi:jcp:02}.
They employed monomer-resolved Molecular Dynamics (MD) simulations and analytical
theories and found an ultra-soft, bounded, density-dependent effective interaction
governed by the entropic repulsions of counterions trapped in the interior of
the stars. The good agreement between simulations and theory even allowed them
to put forward analytic expressions for the full pair potential at arbitrary star
separations. The effective potential has a weak density-dependence,
which however disappears when 
the star density exceeds its overlap value $\rho_{\rm s}^*$.
In this case,
all counterions are absorbed within the stars, whose bare charges are 
therefore completely compensated. 
Thus, the effective potential vanishes identically
for centre-to-centre distances $r>\sigma_{\rm s}$. For overlapping distances
$r\leq\sigma_{\rm s}$, there is no dependence on the concentration anymore and 
only the trapped counterions' entropy contributes to the star--star interaction,
for this reason reading for $r \leq \sigma_{\rm s} \equiv q\sigma_{\rm c}$
as \cite{jusufi:prl:02,jusufi:jcp:02,hoffmann:jpcm:03,hoffmann:jcp:04}:
\begin{equation}
\fl
\frac{\beta V_{\rm ss}(r)}{2N_2}  = 
\ln\left\{\frac{N_2}{2\pi\left[1+\frac{r}{q\sigma_{\rm c}}
\left(1-\ln\left(\frac{r}{q\sigma_{\rm c}}\right)\right)\right]}\right\}
+\frac{\frac{r}{q\sigma_{\rm c}}\ln^2\left(\frac{r}{q\sigma_{\rm c}}\right)}
{1+\frac{r}{q\sigma_{\rm c}}\left(1-\ln\left(\frac{r}{q\sigma_{\rm c}}\right)\right)}
- \ln\left(\frac{N_2}{4\pi}\right).
\label{vss:eq}
\end{equation}
In eq.\ (\ref{vss:eq}) above, $N_2$ is the number of spherically
trapped counterions of a single star. It does not coincide with the number
of released counterions, $M$, because the number $N_1$ of 
Manning-condensed counterions \cite{manning:jcp:69} does not contribute
to the effective interaction and must be excluded: thus $N_2 = M - N_1$.
Extensive simulations \cite{jusufi:prl:02,jusufi:jcp:02,konieczny:jcp:06,konieczny:molsym} have shown 
that the relative population of counterions in the two possible states 
is essentially independent of $r$.  Thus, we fix $N_1$ 
to the value measured in MD simulations
made during the investigation of PE-stars in planar 
confinement \cite{konieczny:jcp:06}. The fraction
$N_1/M$ typically grows with increasing $\alpha$ and covers ranges 
between $30\%$ and $50\%$. 

Clearly, the interaction $V_{\rm ss}(r)$ of
eq.\ (\ref{vss:eq}) vanishes, along with its first derivative with
respect to $r$, at $r = q\sigma_{\rm c}$, guaranteeing the
smooth transition to the region $r > q\sigma_{\rm c}$, in which
$V_{\rm ss}(r) = 0$. The latter feature is, strictly speaking, valid
only for star densities exceeding the overlap value $\rho_{\rm s}^{*}$
\cite{hoffmann:jcp:04}. For $\rho_{\rm s}<\rho_{\rm s}^*$, a Yukawa
tail exists, emerging from the Coulomb interaction between the 
non-neutralised PE-stars and screened by the free counterions. 
For the purposes of simplicity, we ignore this small contribution,
because the number of released counterions from multiarm PE-stars 
constitutes, at all densities, a tiny fraction of the total number 
of counterions $M$ \cite{jusufi:jcp:02}, as 
confirmed by the very small values of
experimentally measured osmotic coefficients 
of PE-star solutions \cite{jusufi:cps:04}.

\subsection{The cross interaction}
\label{sec:eff_pot2}

In order to complete the set of effective pair potentials needed to describe the
binary mixture within the framework of a full two-component picture, we have to specify
the colloid--PE-star cross interaction. Thereby, we proceed along the lines of Ref.\
\cite{jusufi:jpcm:13} to derive the desired potential for small $q$-values
based on results for the effective repulsion in the case where a PE-star is brought 
within a distance $z$ from a hard, flat wall \cite{konieczny:jcp:06,konieczny:molsym}. 

\begin{figure}
  \begin{center}
    \includegraphics[height=6cm,clip]{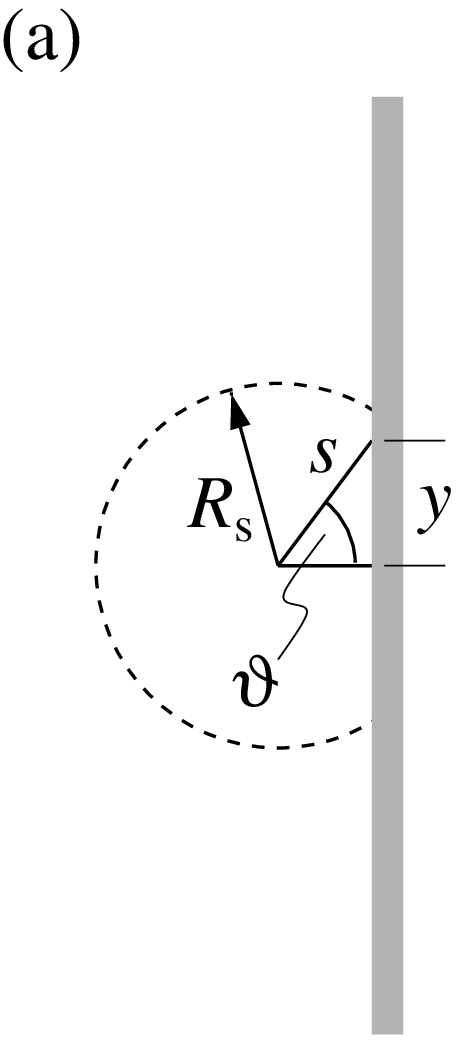}
    \includegraphics[height=6cm,clip]{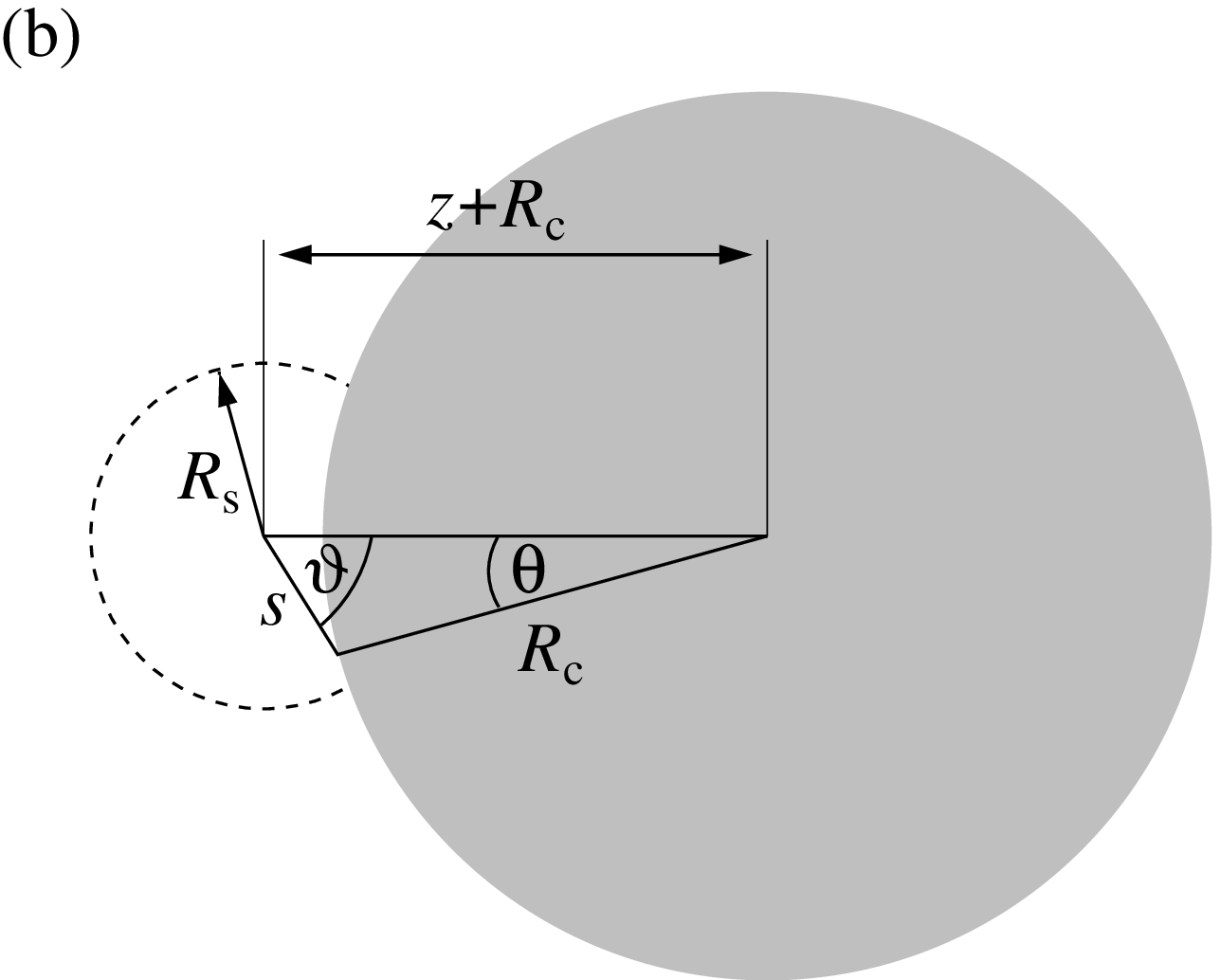}
    \caption{\label{fig:geometry}
    PE-star (smaller, dashed sphere) interacting with (a) a planar wall or
    (b) a hard colloidal particle (bigger, solid-gray sphere).
    }
  \end{center}
\end{figure}

To begin with, let $V_{\rm sw}(z)$ be the star--wall interaction and 
$F_{\rm sw}(z)=-\partial V_{\rm sw}(z)/\partial z$ the corresponding force
for a PE-star with all its counterions absorbed, i.e., for densities $\rho_{\rm s}$
beyond the overlap density (see also previous Sec.\ \ref{sec:eff_pot1}). Then, 
for the geometry shown in fig.\ \ref{fig:geometry}(a), the force is related to
the osmotic pressure $\Pi(s)$ exerted by the star on the surface of the wall via
integration of the normal component of the latter along the area of 
contact \cite{pincus:macromol:24}:
\begin{equation}
F_{\rm sw}(z)=2\pi\int_0^\infty{\rm d}y\,y\,\Pi(s)\cos\vartheta
=2\pi z\int_z^\infty{\rm d}s\,\Pi(s).\label{eq:cross1}
\end{equation}
Using the above eq.\ \ref{eq:cross1}, we can directly obtain the functional form
for the osmotic pressure $\Pi(z)$, provided that the functional form for the
star--wall force $F_{\rm sw}(z)$ is known:
\begin{equation}
\label{eq:cross2}
\Pi(z)=-\frac{1}{2\pi}\frac{\rm d}{{\rm d}z}\left(\frac{F_{\rm sw}(z)}{z}\right).
\end{equation}

The same ideas can in principle be applied for a PE-star in the vicinity of a 
spherical, hard colloid, i.e., a hard sphere. Again, integrating the osmotic 
pressure along the area of contact between both objects yields the force acting
on the centres of the mesoscopic particles. 
Pursuant to the geometry of the problem,
see fig.\ \ref{fig:geometry}(b), and paying regard to the underlying symmetry, 
we get as result for the colloid--PE-star cross force $F_{\rm cs}^*(z)$:
\begin{equation}
F_{\rm cs}^*(z)=\frac{\pi\sigma_{\rm c}^2}{2}\int_0^{\theta_{\rm max}}
{\rm d}\theta\,\sin\theta\,\Pi(s)\cos\vartheta.
\end{equation}
Here, the upper integration boundary $\theta_{\rm max}$ can be 
acquired by the condition that $\Pi(s)$ must vanish identically for all 
$\theta>\theta_{\rm max}$. 
It is possible to eliminate the polar angles $\vartheta$ and $\theta$ emanating from the
centres of the PE-star and the colloid, respectively, in favor of the distance $s$
between the star centre and the point on the colloid's surface that is determined 
by the aforementioned angles. In doing so, we use geometrical relations evident 
from the sketch in fig.\ \ref{fig:geometry}(b), and finally obtain:
\begin{equation}
\label{eq:cross3}
F_{\rm cs}^*(z)=\frac{\pi\sigma_{\rm c}}{2\left(\sigma_{\rm c}+2z\right)^2}
\int_z^{s_{\rm max}}{{\rm d}s\,\left[\left(\sigma_{\rm c}+2z\right)^2
-\sigma_{\rm c}^2+4s^2\right]\Pi(s)}.
\end{equation}
Again, we may obtain the maximum integration distance $s_{\rm max}$ (without any 
need to calculate $\theta_{\rm max}$ before) simply by demanding that $\Pi(s)$ 
must be equal
to zero for all $s>s_{\rm max}$. For such values of $s$, the integrand as a whole
obviously vanishes and there are no contributions to the result of the integration 
anymore. Presumed the functional form for the osmotic pressure is known, such 
identification of $s_{\rm max}$ is rather easily feasible.

\begin{figure}[t]
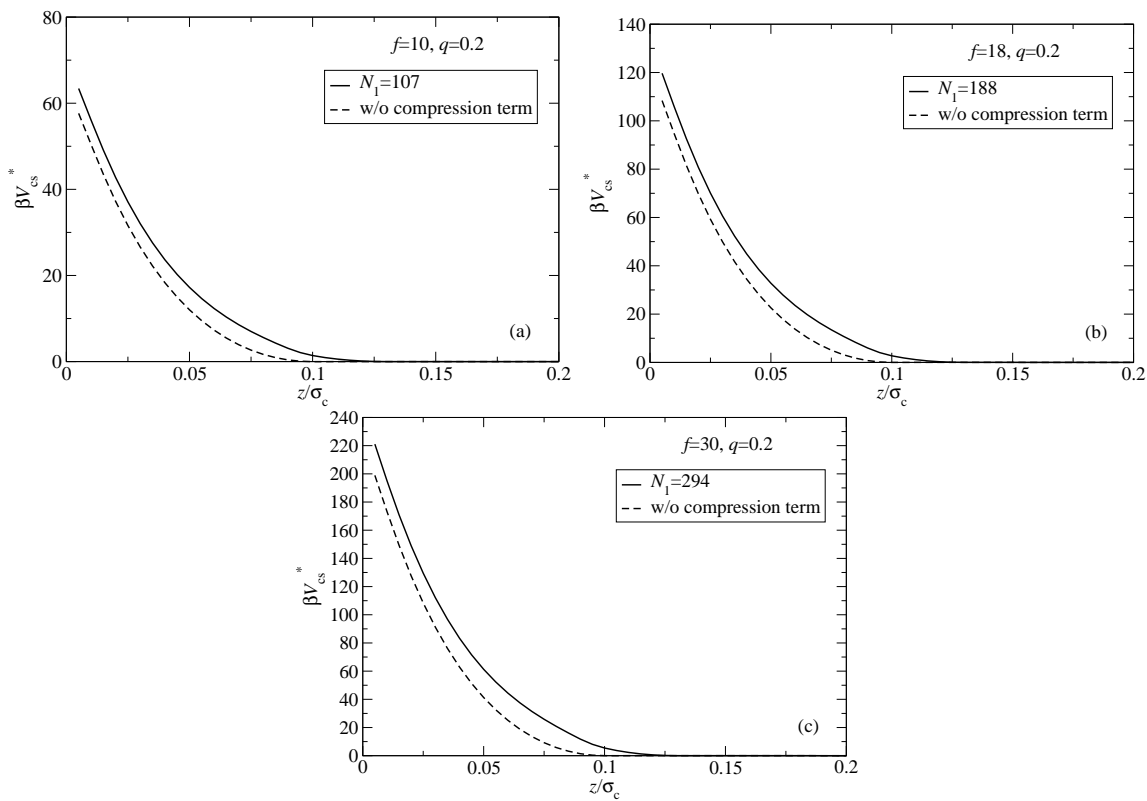

  \begin{center}
    \includegraphics[width=7.5cm,clip]{./figure2a.eps}
    \includegraphics[width=7.5cm,clip]{./figure2b.eps}
    \includegraphics[width=7.5cm,clip]{./figure2c.eps}
    \caption{\label{fig:Vsc*}
    Effective colloid--PE-star potentials with and without provision for 
    compression effects as a function of the centre-to-surface
    distance $z$. Here, we have chosen the parameters $q=0.2$ and 
    (a) $f=10$, (b) $f=18$, 
    and (c) $f=30$. In the legend boxes, the numbers of condensed counterions 
    $N_1$ used as fit parameters in Ref.\ \cite{konieczny:jcp:06} are specified
    for sake of completeness.
    }
\end{center}
\end{figure}

Since we want to consider small values $q\leq 0.3$ of the size ratio only, the
stars discern the colloidal surface to be rather weakly bent compared to
a flat wall, i.e., the radius of curvature is large in terms of the star 
diameter $\sigma_{\rm s}$. 
Therefore, it is a reasonable approximation to assume that
the osmotic pressure remains almost unchanged with respect to the situation
where a PE-star is brought in contact with a planar wall. Consequently, we may
combine eqs.\ (\ref{eq:cross2}) and (\ref{eq:cross3}) to obtain a sound estimate
for the effective force $F_{\rm cs}^*(z)$ as a function 
of distance of the star centre and the colloid's surface. 
Note that in our special case $s_{\rm max}$ is of the order of the star radius 
$R_{\rm s}$. This fact becomes evident from eq.\ (\ref{eq:cross2}) if one takes 
into account that the typical range for the star--wall force $F_{\rm sw}(z)$
is also approximately $R_{\rm s}$ or at the utmost slightly bigger due to
effects of a chain compression at the hard wall (see below) 
\cite{konieczny:jcp:06}. Clearly, the corresponding potential is received by 
a simple, one-dimensional integration:
\begin{equation}
\label{eq:Vsc*}
V_{\rm cs}^*(z)=\int_\infty^z{\rm d}z'\,F_{\rm cs}^*(z').
\end{equation}
In fig.\ \ref{fig:Vsc*} we show the shape of $V_{\rm cs}^*(z)$  for
$q=0.2$ and different values of the stars' functionality $f$. In order to
demonstrate the importance of the so-called compression term adding
to the star--wall force $F_{\rm sw}$ besides electrostatic-entropic contributions
\cite{konieczny:jcp:06}, we additionally included colloid--star potentials
calculated on the basis of the electrostatic and entropic star--wall forces 
alone. Since there are striking deviations, we can clearly expect
such devolved compression effects to influence the phase behaviour of the
mixture.

Finally, we need to express
the effective potential as a function of the particles' 
centre-to-centre separation $r$ instead of the centre-to-surface distance $z$.
Thereby, we have to take into account that the star centre is strictly forbidden 
to penetrate the volume of the colloid. Thus, the total cross interaction features 
a hard core plus the soft, purely repulsive tail as obtained from the above 
eq.\ (\ref{eq:Vsc*}) and finally writes as:
\begin{equation}
V_{\rm cs}(r)=
\left\{
\begin{array}{ll}
\infty & r\leq\sigma_{\rm c}/2\\
V_{\rm cs}^*(r-\sigma_{\rm c}/2) & {\rm else}.
\end{array}
\right.
\end{equation}

\section{Determination of the structure and thermodynamics of the mixture}
\label{sec:iet}

In this section, we describe the basic principles of liquid integral equation 
theory for binary mixtures\footnote{
A further generalization of the theoretical approach from $\nu=2$ to $\nu>2$ 
components in the mixture is straightforward. But since we are only interested 
in binary systems within the framework of this paper, we limit ourselves to
that special case in order to keep the delineation as concise as possible.}
and how to subsequently access the thermodynamics of the system. In general,
the pair structure of the system at hand (and analogously any other two-component 
system) is fully described by three independent
total correlation functions $h_{ij}(r)$ with $i,j={\rm c,s}$. Hereby, we already
allowed for the symmetry with respect to exchange of the indices, i.e.,
$h_{ij}(r)=h_{ji}(r)$. Closely related to the total correlation functions
are the so-called direct correlation functions (dcf's) $c_{ij}(r)$. 
Following the same symmetry argument again, there exist only three independent 
dcf's. In what follows, we will denote the Fourier transforms of $h_{ij}(r)$
and $c_{ij}(r)$ as $\tilde h_{ij}(r)$ and $\tilde c_{ij}(r)$, respectively.

The above-mentioned connection between the total and direct correlation functions 
is quantitatively incorporated via the multicomponent generalization of the 
well-known and commonly used Ornstein-Zernike (OZ) relation, which in its Fourier 
space representation reads as \cite{hansen:book,hansen:arpc:51, lebowitz:jcp}:
\begin{equation}
\label{eq:oz1}
{\bf\tilde H}(k)={\bf\tilde C}(k)+{\bf\tilde C}(k)\cdot{\bf D}
\cdot{\bf\tilde H}(k).
\end{equation}
Here, ${\bf\tilde H}(k)$ and ${\bf\tilde C}(k)$ are symmetric $(2\times 2)$ 
matrices whose elements are constituted by the total and direct correlation 
functions, respectively, and ${\bf D}$ is a diagonal $(2 \times 2)$ matrix 
containing the partial densities characterising the composition of the 
system under investigation, i.e.,
\begin{equation}
\left[{\bf\tilde H}(k)\right]_{ij}=\tilde h_{ij}(k),\;
\end{equation}
\begin{equation}
\left[{\bf\tilde C}(k)\right]_{ij}=\tilde c_{ij}(k),
\end{equation}
\begin{equation}
\left[{\bf D}\right]_{ij}=\rho_i\delta_{ij}.
\end{equation}
Evidently, eq.\ (\ref{eq:oz1}) can be rewritten yielding 
the equivalent matrix relation
\begin{equation}
{\bf\tilde H}(k)=\left[{\bf 1}-{\bf\tilde C}(k)\cdot{\bf D}
\right]^{-1}\cdot{\bf\tilde C}(k),
\end{equation}
with the identity matrix ${\bf 1}$ and the matrix 
inverse $\left[{\bf 1}-{\bf\tilde C}(k)\cdot{\bf D}
\right]^{-1}$.
Defining $\Delta(k)\equiv\rho_{\rm s}\rho_{\rm c}
[\tilde c_{\rm ss}(k)\tilde c_{\rm cc}(k)
-\tilde c_{\rm cs}^2(k)]$ and $E(k)\equiv\rho_{\rm s}\tilde c_{\rm ss}(k)
+\rho_{\rm c}\tilde c_{\rm cc}(k)$
and returning to a component-by-component notation, the latter can consistently
be expressed in the following fashion:
\begin{equation}
\label{eq:oz2}
\tilde h_{ij}(k)=\frac{\tilde c_{ij}(k)-\rho_i^{-1}\cdot
\Delta(k)\cdot\delta_{ij}}{1+\Delta(k)-E(k)}.
\end{equation}

The  
linear algebraic system of eq.\ (\ref{eq:oz2}), 
provides three independent equations 
coupling six yet unknown functions
$\tilde h_{ij}(k)$ and $\tilde c_{ij}(k)$. In order to completely determine that 
set of functions, we therefore need to supply three additional relations to close
and subsequently solve the system of equations. There are several popular
choices for these so-called closures, e.g., the Percus-Yevick (PY) or 
hypernetted-chain (HNC) approximations in their respective two-component 
generalizations. While the former is known to generate reliable results for 
short-range interactions, the latter furnishes very accurate
estimates for the pair structure in case of long-ranged, soft potentials.
Neither the PY
nor the HNC closure are thermodynamically consistent,
however, and in our case this is a crucial factor, since we are
interested in the calculation of phase boundaries, which should not depend
on the route chosen to calculate the free energies. Thus, we resort to
the Rogers-Young (RY) closure \cite{rogers:pra:84}, 
in which thermodynamic consistency
can be enforced. In its multicomponent version the 
RY-closure reads as:
\begin{equation}
\label{eq:ry}
g_{ij}(r)=\exp\left[-\beta V_{ij}\right]\cdot\left\{
1+\frac{\exp\left[\chi_{ij}(r)f_{ij}(r)\right]-1}
{f_{ij}(r)}\right\},
\end{equation}
where $g_{ij}(r)=h_{ij}(r)+1$ are the so-called radial distribution functions and
we introduced new auxiliary functions $\chi_{ij}(r)=h_{ij}(r)-c_{ij}(r)$.
$V_{ij}(r)$ refers to the pair interactions between species $i$ and $j$ as 
presented in sec.\ \ref{sec:eff_pot}. It may be again emphasised that the main 
benefit we gain from using the modified relation (\ref{eq:ry}) is closely related to 
the hybrid character of the latter. Due to the fact that any closure constitutes 
an approximation, we in general obtain different results for the partial and 
total isothermal compressibilities as calculated via either the virial or the 
fluctuation route (see below), as already mentioned above. But the three mixing 
functions emerging in eq.\ (\ref{eq:ry}) above and given by 
\begin{equation}
\label{eq:ry2}
f_{ij}(r)=1-\exp\left[-\zeta_{ij}r\right],
\end{equation}
with $\zeta_{ij}$ being the so-called self-consistency parameters, now allow us 
to address this problem and to appropriately match the isothermal compressibilities. 
Since it is
sufficient to apply a single consistency condition only, namely the requirement 
of equality of the system's total virial and fluctuation isothermal compressibilities, 
the usual approach is to employ just one individual parameter 
$\zeta_{ij}=\zeta$ for all components. Hence, only a single mixing function 
$f_{ij}(r)=f(r)$ remains. However, multi-parameter versions of the RY closure have 
nevertheless also been proposed some years ago \cite{biben:jpcm:91}, accordingly 
demanding the equality of all the partial compressibilities. It is
easy to check that for $\zeta=0$ and $\zeta=\infty$ the multicomponent PY and 
HNC closures, respectively, are recovered from eq.\ (\ref{eq:ry})\footnote{
When using the RY closure, the correlation functions obviously, besides their 
inherent density dependence, parametrically depend on the mixing parameter $\zeta$,
i.e., $h_{ij}=h_{ij}(r;\rho_{\rm c},\rho_{\rm s},\zeta)$ and $c_{ij}=c_{ij}(r;\rho_{\rm c},\rho_{\rm s},\zeta)$. The same must obviously hold
for all quantities deduced from these two functions. Note that we will 
nevertheless throughout this paper drop both the $\rho_i$'s and $\zeta$ from 
the respective parameter lists in order not to overcrowd our notation.}.

Now, we have to address in more detail the issue 
of calculating the total isothermal 
compressibility following the different routes.  At first, we
concern ourselves with the virial compressibility $\kappa_T^{\rm v}$. The total pressure 
$P$ of the system at hand, including both ideal and 
excess contributions, takes the 
form \cite{lebowitz:jcp}:
\begin{equation}
\label{eq:pressure}
\beta P=\rho-\frac{2\pi\rho^2}{3}\sum_i\sum_jx_ix_j\int_0^\infty
{\rm d}r\,r^3\,V_{ij}'(r)\,g_{ij}(r),
\end{equation}
with $V_{ij}'(r)=-\partial V_{ij}(r)/\partial r$ being the different pair
potentials' derivatives with respect to the inter-particle distance $r$. Provided
the pressure pursuant to eq.\ (\ref{eq:pressure}) is known, $\kappa_T^{\rm v}$ can 
be obtained by differentiating with respect to the total density $\rho$ while the 
partial concentrations $x_i$ are kept fixed:
\begin{equation}
\rho k_{\rm B}T\kappa_T^{\rm v}=\left[\left.\frac{\partial(\beta P)}
{\partial \rho}\right|_{\{x_i\}}\right]^{-1}.
\end{equation}

In order to evaluate the fluctuation compressibility $\kappa_T^{\rm fl}$, we 
initially introduce the three partial structure factors $S_{ij}(k)$. As the
correlation functions and the radial distribution functions, respectively, they  
also describe the structure of the system:
\begin{equation}
S_{ij}(k)=\delta_{ij}+\sqrt{\rho_i\rho_j}\tilde h_{ij}(k).
\end{equation}
While for the one-component case the compressibility can simply be obtained as
the $(k=0)$-value of the static structure factor, i.e., 
$S(k=0)=\rho k_{\rm B}T\kappa_T^{\rm fl}$, things are a bit more complicated 
for binary (or multicomponent, $\nu>2$) mixtures. Here, in generalization of 
the one-component situation, the compressibility can finally be written using
the following expression \cite{kirkwood:jcp:51,ashcroft:78,likos:jcp:92}:
\begin{equation}
\rho k_{\rm B}T\kappa_T^{\rm fl}=\frac{S_{\rm ss}(0)S_{\rm cc}(0)-S_{\rm cs}^2(0)}
{x_{\rm c}S_{\rm ss}(0)+x_{\rm s}S_{\rm cc}(0)-2\sqrt{x_{\rm s}x_{\rm c}}
S_{\rm cs}^2(0)}.
\end{equation}

Based on the knowledge of the partial correlation functions $h_{ij}(r)$ and 
structure factors $S_{ij}(k)$ as obtained by (numerically) solving the OZ 
relation, eq.\ (\ref{eq:oz1}), and using the RY closure, eq.\ (\ref{eq:ry}), we 
can in principle completely access the thermodynamics of the system at hand. 
In order to calculate the binodal lines, a very convenient quantity to 
consider is the concentration structure factor $S_{\rm con}(k)$. It is a linear 
combination of all the partial structure factors, whereby the corresponding 
pre-factors are determined by the different species' concentrations $x_i$, namely:
\begin{equation}
S_{\rm con}(k)=x_{\rm c}x_{\rm s}^2S_{\rm cc}(k)
+x_{\rm s}x_{\rm c}^2S_{\rm ss}(k)
-2(x_{\rm c}x_{\rm s})^{3/2}S_{\rm cs}(k).
\end{equation}
Now, let $P$ be the total pressure according to the above eq.\ (\ref{eq:pressure}) 
and $g(x_{\rm s},P,T)=G(x_{\rm s},N,P,T)/N$ the Gibbs free energy 
$G(x_{\rm s},N,P,T)$ per particle. Then, the second derivative of the 
former is 
connected to the concentration structure factor $S_{\rm con}(k)$ by means of the 
sum rule \cite{biben:prl:91,bhatia:pre:70,dzubiella:jcp:116}:
\begin{equation}
\label{eq:dgl}
\beta g''(x_{\rm s},P,T)
\equiv\beta\frac{\partial^2g(x_{\rm s},P,T)}{\partial x_{\rm s}^2}
=\frac{1}{S_{\rm con}(0;x_{\rm s})},
\end{equation}
where we have added the concentration $x_{\rm s}$ as a second argument
to $S_{\rm con}(k)$ to emphasise this dependence. This differential
equation has to be integrated along an isobar for any prescribed
value of the pressure $P^*\equiv\beta P\sigma_{\rm c}^3={\rm const}$, 
to obtain the Gibbs free energy from the
structural data, $S_{\rm con}(k=0;x_{\rm s})$.
A detailed analysis of
the limiting behaviour of $g''(x_{\rm s})$ shows a divergence as $1/x_{\rm s}$ for
$x_{\rm s}\rightarrow 0$ and as $1/(1-x_{\rm s})$ for $x_{\rm s}\rightarrow 1$
\cite{dzubiella:jcp:116}. In order to avoid any technical difficulties when
numerically integrating, we a priori split the Gibbs free energy $g(x_{\rm s})$
into a term that arises from its ideal part and a remainder,
which we call excess part\footnote{The `excess' part
$g_{\rm ex}(x_{\rm s})$ in eq.\ (\ref{eq:gibb1}), includes 
a term $\ln(\rho\sigma_{\rm c}^3)$ that arises from the original
ideal part but which does not cause any divergences at the
limits $x_{\rm s}\to 0$ and $x_{\rm s}\to 1$, which we
seek to remove. Thus we readsorb it into a redefined excess
part, which can be integrated without problems.}, $g_{\rm ex}(x_{\rm s})$:
\begin{eqnarray}
\nonumber
\beta g(x_{\rm s})& = & (1-x_{\rm s})\ln(1-x_{\rm s})+x_{\rm s}\ln(x_{\rm s})
\\
& + &
3(1-x_{\rm s})\ln(\Lambda_{\rm c}/\sigma_{\rm c}) 
+ 3x_{\rm s}\ln(\Lambda_{\rm s}/\sigma_{\rm c})
+\beta g_{\rm ex}(x_{\rm s}),
\label{eq:gibb1}
\end{eqnarray}
with the thermal de Broglie wavelengths $\Lambda_{\rm c,s}$ of the
colloids and the stars, respectively.
Taking the second derivative in the above eq.\ (\ref{eq:gibb1}) again, we 
obtain:
\begin{equation}
\label{gexpp:eq}
\beta g''(x_{\rm s})=\frac{1}{x_{\rm s}}+\frac{1}{1-x_{\rm s}}
+g_{\rm ex}''(x_{\rm s}).
\end{equation}
Thus, the ideal part of the Gibbs free energy is 
exclusively responsible for the
appearance of the aforementioned divergences at the integration boundaries and
the modified second-order differential equation 
\begin{equation}
\label{eq:dgl2}
\beta g_{\rm ex}''(x_{\rm s})=\frac{1}{S_{\rm con}(0;x_{\rm s})}
-\left(\frac{1}{x_{\rm s}}+\frac{1}{1-x_{\rm s}}\right)
\end{equation}
for the excess Gibbs free energy alone is obviously 
free of any diverging terms. We 
can therefore easily solve it numerically. Subsequent addition of the analytically 
known ideal term 
$g_{\rm id}(x_{\rm s})=(1-x_{\rm s})\ln(1-x_{\rm s})+x_{\rm s}\ln(x_{\rm s})$
directly yields the total Gibbs free energy per particle that we are interested 
in. The two terms involving the thermal de Broglie wavelength are
linear in $x_{\rm s}$; they only provide a shifting of the chemical
potentials and can be dropped.

Thermodynamic stability requires that $g(x_{\rm s})$ is convex \cite{row:book}.
In case we encounter some $x_{\rm s}$-region where $g''(x_{\rm s}) < 0$
the binary 
mixture features a fluid-fluid demixing transition. In that sense, 
eqs.\ (\ref{eq:dgl}) and (\ref{eq:dgl2}), respectively, 
allow us to investigate the 
thermodynamics and the phase behaviour of the system at hand by providing a tool 
to compute the Gibbs free energy (per particle). The corresponding phase 
boundaries can be calculated using Maxwell's common tangent construction,
which
guarantees that the chemical potentials, $\mu_i$, are
the same between both coexisting phases. Since we are in a situation where we 
moreover fixed the pressure $P^*$ of the mixture and its absolute temperature $T$, 
all conditions for phase coexistence are clearly fulfilled. Concretely,
the common tangent construction amounts to solving the coupled equations
\begin{equation}
\label{ct1:eq}
g'(x_{\rm s}^{\rm I}) = g'(x_{\rm s}^{\rm II})
\end{equation}
and
\begin{equation}
\label{ct2:eq}
g(x_{\rm s}^{\rm I}) - x_{\rm s}^{\rm I}g'(x_{\rm s}^{\rm I}) 
 = g(x_{\rm s}^{\rm {II}}) - x_{\rm s}^{\rm {II}}g'(x_{\rm s}^{\rm {II}})
\end{equation}
for the concentrations $x_{\rm s}^{\rm {I,II}}$ of the coexisting
phases I and II. 

In integrating eq.\ (\ref{gexpp:eq}) above and adding the
ideal terms, one obtains the 
Gibbs free energy per particle, $g(x_{\rm s})$ 
modulo an undetermined linear function $C_1x_{\rm s} + C_0$ with
the constants $C_1$ and $C_0$ to be fixed by appropriate boundary
conditions. As is clear from eqs.\ (\ref{ct1:eq}) and (\ref{ct2:eq})
above, such a linear term is anyway immaterial from the determination
of phase boundaries and, in practice, it can be ignored on the
same grounds that the terms involving the thermal de Broglie 
wavelengths in eq.\ (\ref{eq:gibb1}) have been dropped. Nevertheless,
the constants $C_1$ and $C_0$ can be determined as follows.
Taking into account that the Gibbs free energy $G(N,P,x_{\rm s},T)$ is an 
extensive function but in its 
natural argument list there is only one extensive variable, namely the number 
of particles $N$, Euler's theorem \cite{landau:book} asserts the function $g$ 
to have the form:
\begin{equation}
\label{eq:euler}
g(x_{\rm s})=x_{\rm s}\mu_{\rm s}(x_{\rm s})+(1-x_{\rm s})\mu_{\rm c}(x_{\rm s}).
\end{equation}
For both limiting one-component 
cases, i.e., if no stars ($x_{\rm s}=0$) 
or no colloids ($x_{\rm s}=1$) are present 
in the system, the following relation holds true:
\begin{equation}
\label{eq:boundary}
\beta g=\hat f+\rho\hat f'=\ln(\rho) + \hat f_{\rm ex} + 
\rho \hat f'_{\rm ex},
\end{equation}
where $\hat f=\beta F/N$ denotes the Helmholtz free 
energy per particle and $\hat f'$
its derivative with respect to the density $\rho$; the subscript `ex'
refers to the excess part of $\hat f$.
On the other hand, $\hat f_{\rm ex}'$ 
is connected to the 
excess pressure $P_{\rm ex}$ [as known from eq.\ (\ref{eq:pressure}) above]
via the equation
\begin{equation}
\label{eq:fp}
\hat f'_{\rm ex}=\beta P_{\rm ex}/\rho^2.
\end{equation}
Following eq.\ (\ref{eq:fp}), $\hat f'_{\rm ex}$
can be obtained by integrating the ratio $P_{\rm ex}/\rho^2$
with respect to $\rho$ and applying the additional boundary condition 
$\hat f_{\rm ex}(\rho\rightarrow 0)=0$. Once the Helmholtz free
energies for the pure colloid and PE-star systems are known this way,
the corresponding chemical potentials $\mu_{\rm c}(0)$ and $\mu_{\rm s}(1)$
can be calculated and the conditions
$g(0)=\mu_{\rm c}(0)$ and $g(1)=\mu_{\rm s}(1)$ for any arbitrary pressure $P$
[cf.\ eqs.\ (\ref{eq:euler}) and (\ref{eq:boundary}) above], yield $C_0$ and
$C_1$. Note that for the pure colloidal system, $x_{\rm s}=0$, 
we can avoid the integration route to 
compute the pressure, by using the accurate Carnahan-Starling expressions for 
hard-spheres \cite{carnahan:jcp:69}, which also turn out to be consistent
with the one calculated from the Rogers-Young route, based on
eq.\ (\ref{eq:pressure}) and our results for the 
radial distribution function 
$g(r)$. 

\begin{figure}[t]
  \begin{center}
    \includegraphics[width=11cm,clip,draft=false]{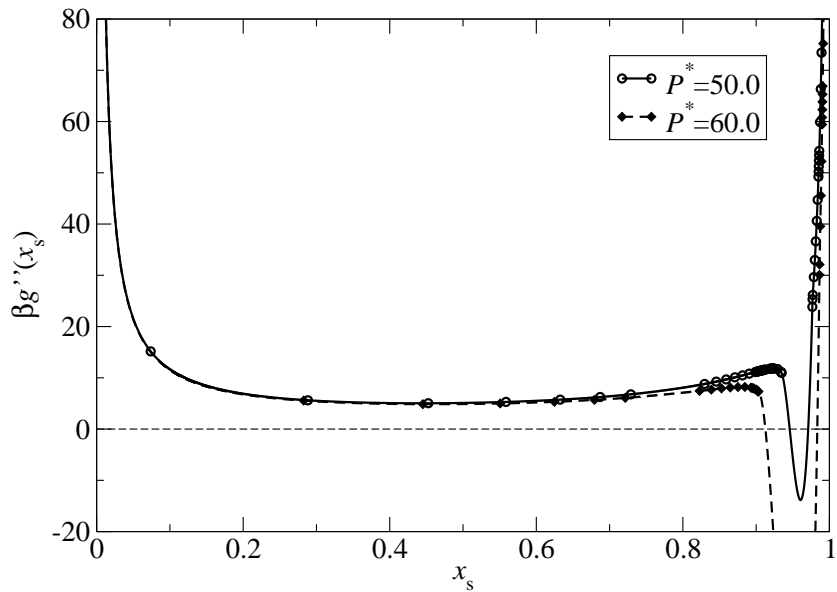}
    \caption{\label{fig:interpolation}
    Examples of the second derivative of the Gibbs free energy per particle,
    $g''(x_{\rm s})$, plotted against the star concentration $x_{\rm s}$ for
    stars with functionality $f=30$, 
    PE-star--colloid size ratio $q=0.2$, and two different pressures 
    $P^*=\beta P\sigma^3_{\rm c}$. Symbols are calculated from the OZ relation, lines
    were obtained by cubic spline interpolation. Note that the $x_{\rm s}$-interval 
    where we are not able to numerically solve the integral equations grows distinctly
    upon increasing the pressure, i.e., as we move away from the critical point.
    Moreover, the plot illustrates the limiting behaviour of $g''(x_{\rm s})$ as
    $1/x_{\rm s}$ for $x_{\rm s}\rightarrow 0$ and as $1/(1-x_{\rm s})$ for 
    the opposite case $x_{\rm s}\rightarrow 1$, respectively. 
    }
\end{center}
\end{figure}

\begin{figure}[t]
  \begin{center}
    \includegraphics[width=11cm,clip]{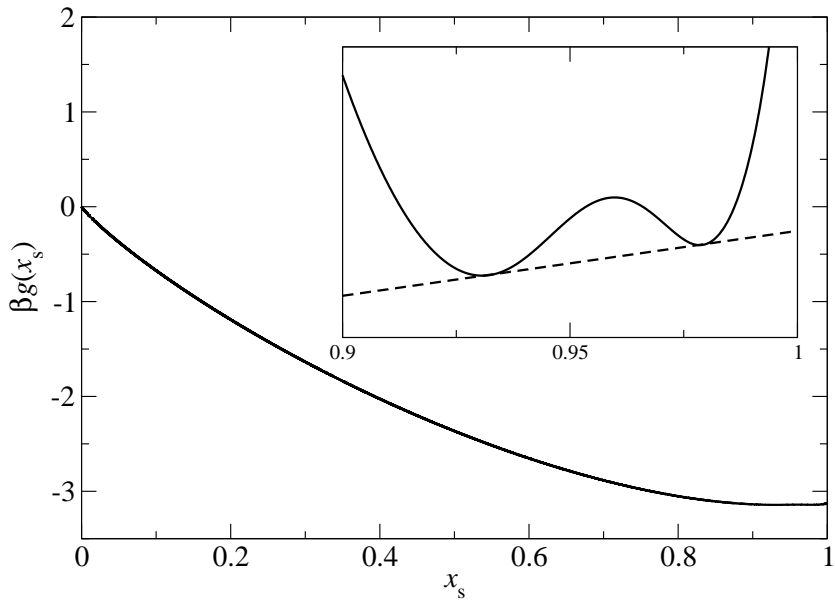}
    \caption{\label{fig:gibb}
    Gibbs free energy per particle $g(x_{\rm s})$ vs.\ the star concentration
    $x_{\rm s}$, plotted for $f=30$, $q=0.2$, and 
    $P^*=50.0$. The curve was obtained via integrating the interpolated function
    $g''(x_{\rm s})$ twice according to the procedure delineated in the main text, 
    whereby we subtracted an arbitrary linear function afterwards. The inset shows 
    $g(x_{\rm s})$ with an differently scaled $x_{\rm s}$-axis in order to highlight 
    the concave parts of the function and to show Maxwell's common tangent construction.
    }
\end{center}
\end{figure}

Note that, when crossing the spinodal line in the density plane, the long
wavelength limits of the partial structure factors, $S_{ij}(k\rightarrow 0)$, take
non-physical values. This behaviour expresses the system's physical instability 
against a possible fluid-fluid phase separation. Thereto, it is not feasible to 
(numerically) solve the integral equations anymore once we reached the spinodal;
in fact, integral equation theories themselves break down before the
spinodal is reached, yet after the binodal \cite{andy:02}. 
Consequently, depending on the total pressure and above a certain threshold value 
of the same, $P>P_{\rm thr}$, the concentration structure factor 
$S_{\rm con}(0;x_{\rm s})$ is unknown over some interval $\Delta x_{\rm s}(P)$.
Hence, we need to appropriately 
interpolate $S_{\rm con}(0;x_{\rm s})$ in order to 
obtain the second derivative of the Gibbs free energy per particle for all 
$0\leq x_{\rm s}\leq 1$ and, in this way, to allow for the integration of
the differential equations (\ref{eq:dgl}) or (\ref{eq:dgl2}), respectively.
Along the lines of Ref. \cite{dzubiella:jcp:116}, we perform this necessary 
interpolation using cubic splines. In order to illustrate the whole procedure,
fig.\ \ref{fig:interpolation} shows the
function $g''(x_{\rm s})$ as computed from the OZ equation together with its 
cubic spline interpolation for a representative parameter combination and
two different pressures. Moreover, in fig.\ \ref{fig:gibb}, we plotted the 
corresponding Gibbs free energy $g(x_{\rm s})$ for the lower one of these 
pressures. In addition, the inset depicts Maxwell's common tangent construction
used to compute the star concentrations for the coexisting phases. 

It may be emphasised that the results for the binodal do not depend on the 
concrete interpolation scheme, at least as long as the numerical
methods used to solve the OZ relation are able to precisely reach the spinodal,
i.e., the points where the structure factors diverge for $k\rightarrow 0$. 
Admittedly, this is not always strictly the case since the numerical schemes we
employed to calculate correlation functions and corresponding structure 
factors, respectively, may break down slightly before the spinodal is reached. 
Accordingly, small inaccuracies induced by the interpolation procedure arise which
grow with increasing width of the gap region $\Delta x_{\rm s}(P)$, or to put it 
in other words, with increasing pressure $P$, i.e., if we move away from the critical
point. As long as the aforementioned 
interval where no solution of the integral equations can be found is rather small,
we expect the interpolation to be reliable, while for higher pressures the
received binodals are of more approximate character. Nevertheless, they still
show a very reasonable behaviour. We are going to discuss the results for the
phase diagrams in detail in sec.\ \ref{sec:phase}.

\section{Results}
\label{sec:res}

\subsection{Low colloid-density limit}
\label{sec:depletion}

Based on the radial distribution functions $g(r)$ as obtained by the OZ relation 
closed with the RY closure, eqs.\ (\ref{eq:oz1}) and (\ref{eq:ry}), we may map our 
two-component mixture onto an effective one-component system of the colloids alone.
In doing so, the PE-stars are completely traced out, resulting into an effective 
colloid--colloid interaction where the pure hard-sphere potential is masked by 
additional depletion contributions originating in the presence of the stars and 
the forces they exert on the colloids. To put it in other words, the colloid--PE-star 
interactions cause spatial correlations of the PE-star distribution in the vicinity of 
the colloids, and it is exactly these correlations that determine the resulting shape 
of the depletion potential. Note that the latter in general parametrically depends 
rather on the PE-stars' chemical potential $\mu_{\rm s}$ or, 
equivalently, the density $\rho_{\rm s}^{\rm r}$ of a reservoir of stars at the 
same chemical potential $\mu_{\rm s}^{\rm r}=\mu_{\rm s}$, than on their density 
$\rho_{\rm s}$ in the real system. Hence, it is in principle more convenient to 
switch to a reservoir representation $(\rho_{\rm c},\rho_{\rm s}^{\rm r})$ of the 
partial densities instead of the original system representation 
$(\rho_{\rm c},\rho_{\rm s})$ when considering such effective interactions. Clearly, 
if the colloid density $\rho_{\rm c}$ takes finite values, it must hold 
$\rho_{\rm s}\neq\rho_{\rm s}^{\rm r}$. But since we will consider the limiting
case of low colloid densities $\rho_{\rm c}\rightarrow 0$ only in what follows, we 
have $\rho_{\rm s}^{\rm r}=\rho_{\rm s}$ again, i.e., reservoir and system 
representation of the partial densities coincide. 

Concretely, the 
desired mapping\footnote{Note that the most accurate way to compute 
effective interactions between 
two colloidal particles in the presence of (smaller) 
PE-stars is to employ direct computer simulations \cite{dzubiella:jcp:116,mayer:pre:70,damico:physa:97,allahyarov:prl:98,allahyarov:jpcm:01}. 
Another way to the depletion potential would in principle be offered by Attard's 
so-called superposition approximation (SA) \cite{attard:jcp:89}. But since we want 
to perform the mapping onto an effective one-component system in order to gain some 
qualitative understanding of the physics of our system only but stick to the full 
two-component picture to quantitatively calculate the binodals of the mixture, we 
turn down such alternative methods within the scope of the paper at hand.}
can be achieved by a so-called inversion of the full, two-component results of 
the integral equations in the low colloid-density limit $\rho_{\rm c}\rightarrow 0$ 
\cite{dzubiella:jcp:116,mayer:pre:70,dzubiella:epl,mendez:pre:00,koenig:pre:01}.
It can be shown from diagrammatic expansions in the framework of the theory of 
liquids \cite{hansen:book} that in this limit the pair correlation function for 
any fluid reduces to the Boltzmann factor $g(r)=\exp[-\beta v(r)]$. Here, $v(r)$ 
denotes the pair potential the fluid's constituent particles interact
by. According to this relation, the effective colloid--colloid potential 
$V_{\rm eff}(r)$, depending parametrically on both the partial colloid and star 
densities $\rho_{\rm c}$ and $\rho_{\rm s}^{\rm r}=\rho_{\rm s}$, is obtained 
as follows:
\begin{equation}
\label{eq:inversion}
\beta V_{\rm eff}(r)=\lim_{\rho_{\rm c}\rightarrow 0}\ln\left[
g_{\rm cc}(r;\rho_{\rm c},\rho_{\rm s}^{\rm r})\right].
\end{equation}

\begin{figure}[t]
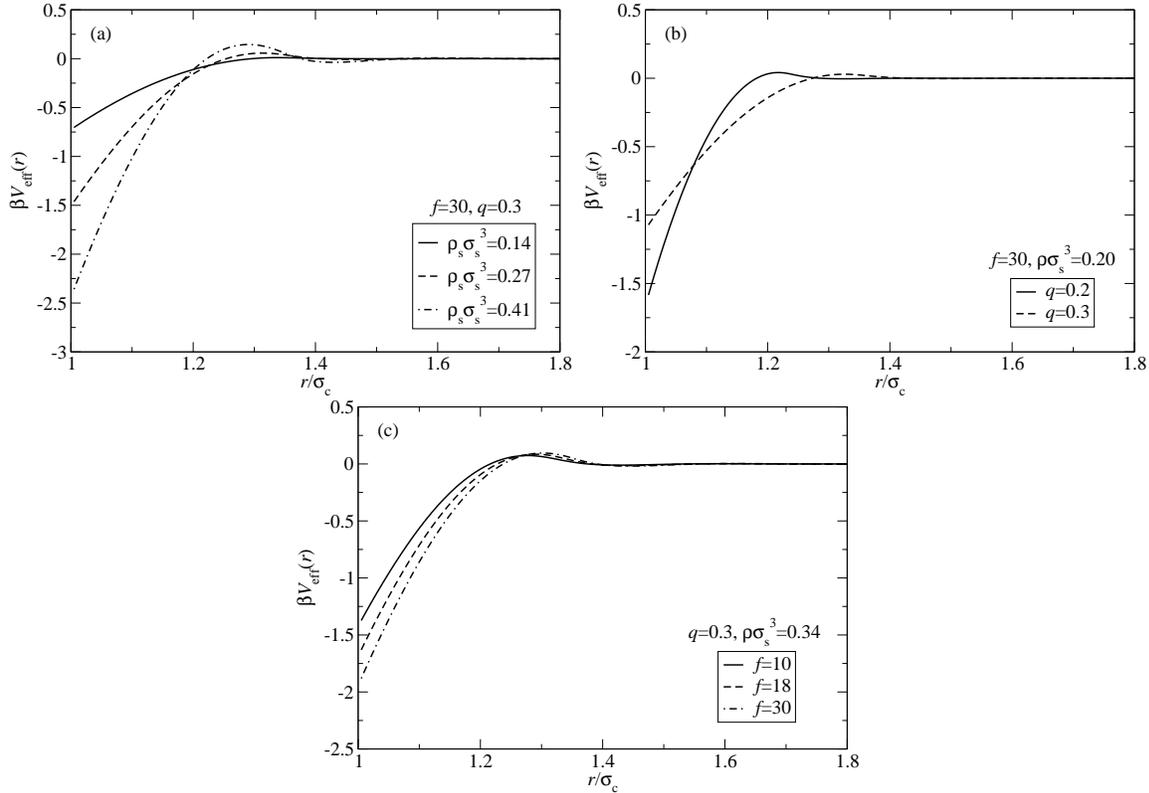

  \begin{center}
    \includegraphics[width=7.5cm,clip]{./figure5a.eps}
    \includegraphics[width=7.5cm,clip]{./figure5b.eps}
    \includegraphics[width=7.5cm,clip]{./figure5c.eps}
    \caption{\label{fig:veff}
    Effective colloid--colloid depletion potentials $V_{\rm eff}(r)$ as 
    obtained by an inversion of the OZ relation. For details concerning the 
    procedure, see main text. We have investigated the influence of (a) the partial
    star density $\rho_{\rm s}^{\rm r}=\rho_{\rm s}$, 
    (b) the PE-star--colloid size ratio $q$, 
    and (c) the stars' functionality $f$
    on the functional form of the interaction potential. It is evident from the  
    plots that the presence of the stars induces an attraction between 
    the colloids in addition to their bare hard-sphere repulsion which takes
    over for distances $r\leq\sigma_{\rm c}$.
    }
\end{center}
\end{figure}

Fig.\ \ref{fig:veff} shows examples for the effective colloid--colloid 
interaction $V_{\rm eff}(r)$ for different functionalities $f$ of the stars,
partial star densities $\rho_{\rm s}^{\rm r}=\rho_{\rm s}$, and PE-star--colloid 
size ratios $q$. As one can see from the plots, for distances $r>\sigma_{\rm c}$ 
the resulting depletion interaction mediated by the stars is attractive and features 
a slightly oscillating behaviour, while for inter-particle separations 
$r\leq\sigma_{\rm c}$ the bare hard-sphere repulsion remains. 
In particular, fig.\ \ref{fig:veff}(a) illustrates that the addition of 
PE-stars to the mixture results in both a significant increase of the depth of the 
attractive potential well and a further enhancement of the aforementioned 
oscillations but does in no way affect the range of the attraction. 
As can be read off
from fig.\ \ref{fig:veff}(b), the latter is determined by the size ratio $q$ alone and
grows linearly with the diameter of the stars. Furthermore, there is a measurable,
indeed weak, dependence of the interaction strength on the functionality of 
the PE-stars: The higher the arm number $f$ gets the stronger becomes the effective 
attraction between two colloids, cf.\ fig. \ref{fig:veff}(c). 
All these trends are in
perfect agreement with the common understanding of the physical mechanisms leading 
to the appearance of such an effective attraction: Due to a depletion of the PE-stars 
in the spatial region between a pair of colloids and dependent on the colloids' mutual 
distance, they are hit asymmetrically by the stars from the inside and the outside. 
Consequently, the unbalanced osmotic pressure pushes the colloids together. Clearly, 
the absolute value of this force must grow when increasing the star density 
$\rho_{\rm s}$, simply because there are more collisions between PE-stars and colloids. 
For higher functionalities $f$, the colloid--PE-star
cross interaction becomes more repulsive (see sec.\ \ref{sec:eff_pot2}), i.e., the
stars push the colloids harder, thus also leading to a strengthened effective 
colloid--colloid attraction. And finally, the PE-stars' diameter determines 
whether or not they fit into the spatial region between a pair of colloids for 
a given distance of the two. Hence, the size ratio $q$ controls if the stars are 
expelled from the said region of space, or to put it in other words, for what scope
of inter-colloidal separations depletion actually takes place. Accordingly, the range 
of the effective force can be altered by changing $q$.

The occurrence of oscillations of the effective potential $V_{\rm eff}$ obviously 
means that the attractive minimum is followed by a repulsive barrier whose 
height is set by the concentration of PE-stars in the mixture, see above. In particular,
it grows upon addition of stars to the system and such behaviour could in case of 
distinctly high and broad maxima in principle lead to micro-phase separation, i.e.,
cluster formation 
\cite{mossa:langmuir:04,sciortino:prl:04,sear:jcp:99,imperio:jcp:06,imperio:jpcm:04}. 
But for the physical system we examine and the range of parameters we investigate, 
the barrier remains anyway rather low and narrow. Micro-phase separation is therefore 
not likely to happen. Instead, the type of effective colloid--colloid attractions at
hand, i.e., an attractive potential valley together with a nearly vanishing or at least 
less-pronounced repulsive barrier, forces the system to develop long-range fluctuations 
upon an increase of the PE-star concentration, consequently favoring the possibility 
of a fluid--fluid demixing transition of the two-component mixture. Such behaviour is
frequently observed in, e.g.,  
colloid--polymer mixtures \cite{vink:pre:05,loverso:jpcm:05,loverso:pre:06}. Thus, when 
considering the phase behaviour of our system by calculating its binodals, we expect 
to find evidence for macro-phase separation. This supposition will be endorsed by the 
results of the following section, too.

\subsection{Structure of the mixture}
\label{struc:sec}

Before switching over to a presentation of the demixing binodals as obtained via 
the procedure described in detail in Sec.\ \ref{sec:iet} of this paper, i.e.,
initially calculating the Gibbs free energy $g(x_{\rm s})$ with both the temperature 
$T$ and the pressure $P$ kept fixed and subsequently identifying the sought-for coexisting 
fluid phases using Maxwell's common tangent construction for the concave parts of that function
(see, in particular, figs.\ \ref{fig:interpolation} and \ref{fig:gibb}), it is useful to 
study partial pair correlation functions $g_{ij}(r)$ and corresponding structure factors 
$S_{ij}(k)$ ($i,j={\rm c,s}$) first. Since these quantities completely describe the pair 
structure of the system, we are able to gain detailed insight into the physics and phase 
behaviour of the mixture and to discover, in addition to the findings of the previous Sec.\ 
\ref{sec:depletion}, more evidence that it is reasonable to expect an mixing--demixing 
transition.

\begin{figure}[t]
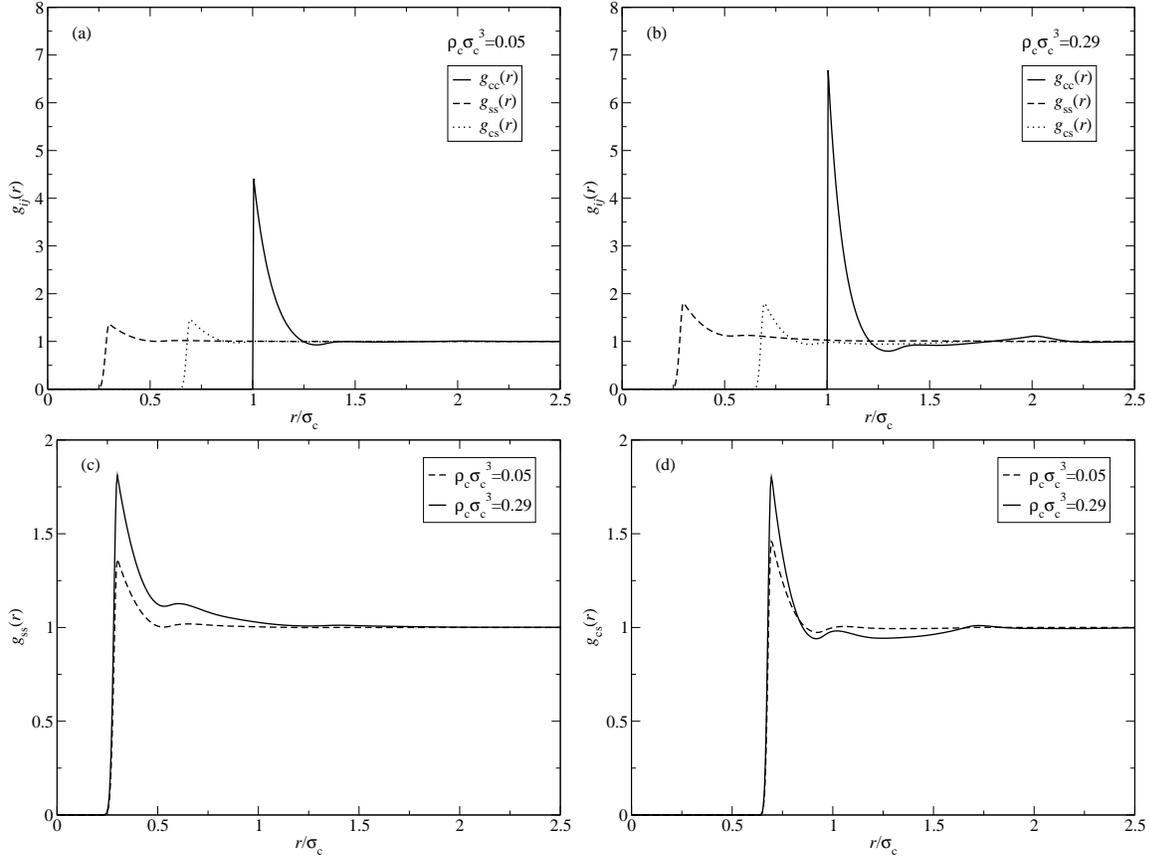

  \begin{center}
    \includegraphics[width=7.5cm,clip]{./figure6a.eps}
    \includegraphics[width=7.5cm,clip]{./figure6b.eps}
    \includegraphics[width=7.5cm,clip]{./figure6c.eps}
    \includegraphics[width=7.5cm,clip]{./figure6d.eps}
    \caption{\label{fig:structure1}
    Partial radial distribution functions $g_{ij}(r)$ ($i={\rm c,s}$) for PE-stars 
    with $f=30$ arms, star--colloid size ratio $q=0.3$, fixed PE-star density 
    $\rho_{\rm s}\sigma_{\rm s}^3=0.27$, and colloidal densities of (a) 
    $\rho_{\rm c}\sigma_{\rm c}^3=0.05$ or (b) $\rho_{\rm c}\sigma_{\rm c}^3=0.29$. 
    For the same value of the stars' partial density, the remaining two parts 
    of the figure illustrate the detailed shape and the $\rho_{\rm c}$-dependence of 
    (c) the star--star correlation function $g_{\rm ss}(r)$ and (d) the 
    cross-correlation function $g_{\rm cs}(r)$. For an in-depth discussion and 
    interpretation of the results, we refer the reader to the main text.
    }
\end{center}
\end{figure}

Fig.\ \ref{fig:structure1} shows the partial 
radial distribution functions $g_{ij}(r)$
for typical parameters, namely a colloid--PE-star mixture with a size 
ratio of $q=0.3$ and the PE-stars having $f=30$ arms each. We show results for different 
mixture compositions, i.e., varying partial densities for both species as indicated in 
the plots. Figs.\ \ref{fig:structure1}(a) and (b), on the one hand, depict the 
decisive length scales of the problem or, equivalently, the typical ranges of 
the underlying pair potentials as set by the the sizes of colloids and PE-stars, 
respectively. The distinct height of the colloid--colloid contact value 
$g_{\rm cc}(\sigma_{\rm c})$ and its further rise upon increasing the PE-star density 
(not shown in our figures) is an obvious manifestation of the mainly attractive 
character of the effective colloid--colloid interactions. In this respect, we again 
refer the reader to Sec.\ \ref{sec:depletion} and, in particular, 
eq.\ (\ref{eq:inversion}) mathematically describing the inversion procedure for the OZ 
relation. On the other hand, when taking a closer look to the whole set of pair correlation 
functions, we find various signs pointing towards the supposable occurrence of a demixing 
transition. The main peaks of both $g_{\rm ss}(r)$ and $g_{\rm cc}(r)$ gain in height 
when adding colloids to the system, while the peak height of the cross-correlation function 
$g_{\rm cs}(r)$ remains essentially the same, see 
figs.\ \ref{fig:structure1}(a), (b) and
(d). In addition, figs.\ \ref{fig:structure1}(c) and (d) 
show an enhancement in the
star--star pair correlations and an concurrent depletion in the colloid--star correlations
for raising colloid densities. The intervals of distances affected are remarkably broad, 
both the range of the enhancement and the depletion are of the order of the colloid 
size, not the much smaller star size. Altogether, these features show the tendency of 
colloids as well as stars to seek spatial proximity of their own species while sort of 
avoiding the other one and we may expect macroscopic regions rich in the one and poor in the 
other species to be formed provided the partial densities, in particular of the colloids, 
are sufficiently high.

\begin{figure}[t]
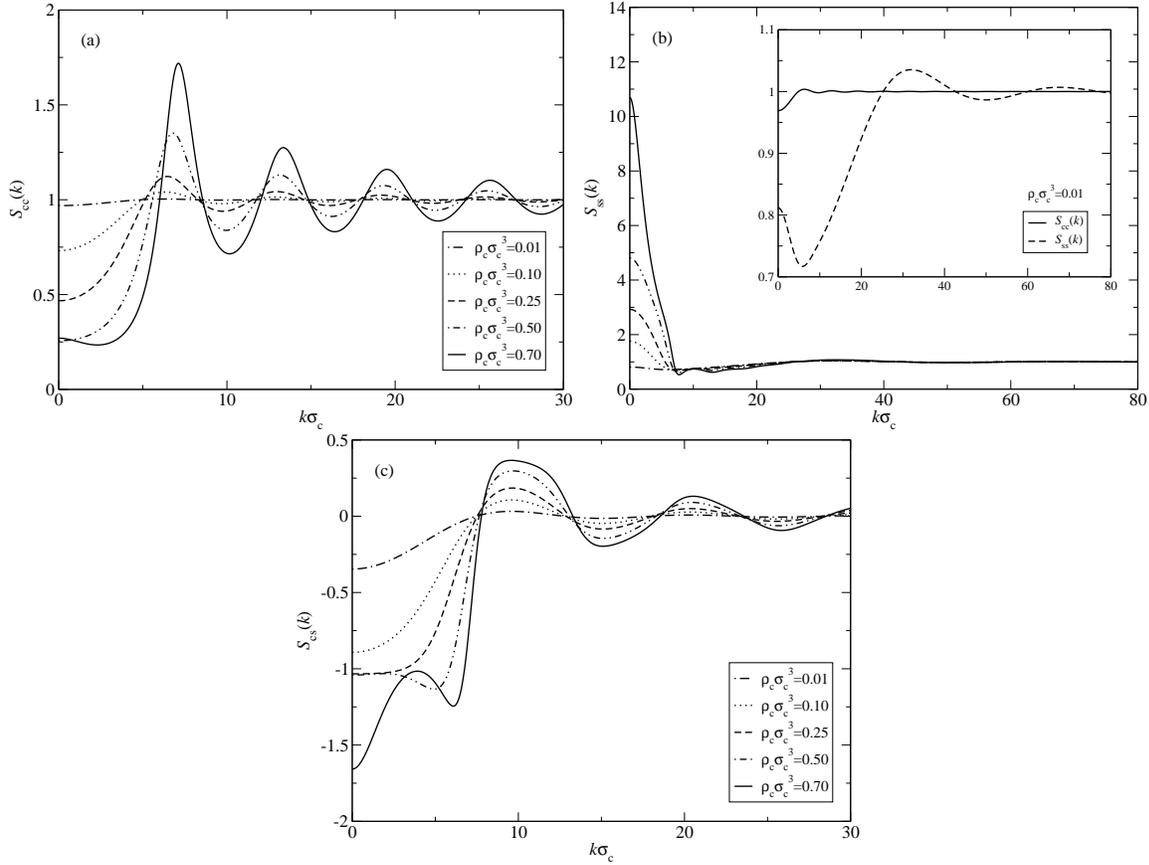

  \begin{center}
    \includegraphics[width=7.5cm,clip]{./figure7a.eps}
    \includegraphics[width=7.5cm,clip]{./figure7b.eps}
    \includegraphics[width=7.5cm,clip]{./figure7c.eps}
    \caption{\label{fig:structure2}
    Examples of the partial structure factors (a) $S_{\rm cc}(k)$, (b) 
    $S_{\rm ss}(k)$,
    and (c) $S_{\rm cs}(k)$ for PE-star functionality $f=18$, size ratio
    $q=0.2$, fixed density of the stars $\rho_{\rm s}\sigma_{\rm s}^3=0.12$,
    and several values of the colloidal density $\rho_{\rm c}\sigma_{\rm c}^3$,
    i.e., different mixture compositions. Please note that the line styles in
    the main plot of part (b) refer to the same parameters as explained in the 
    legends of parts (a) and (c), respectively. The inset in (b) addresses a comparison 
    between the colloid--colloid and the star--star structure factors for the 
    aforementioned star density and a typical value of the colloid 
    density (indicated in the plot) and thereby illustrates the huge difference in 
    the structural length scales of the two species.
    }
\end{center}
\end{figure}

\begin{figure}[t]
  \begin{center}
    \includegraphics[width=11cm,clip]{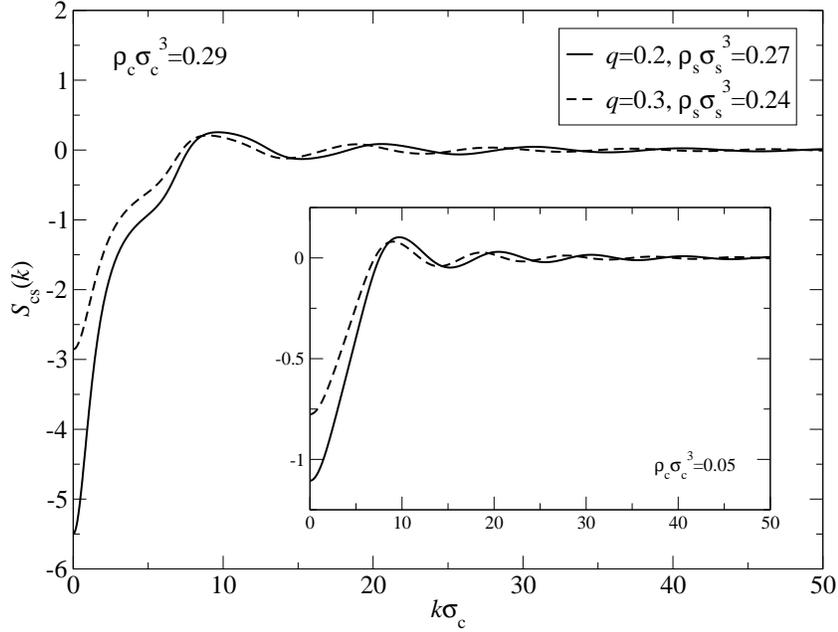}
    \caption{\label{fig:structure3}
    Comparison of the cross structure factors $S_{\rm cs}(k)$ for $f=30$ and the two 
    different size ratios investigated, $q=0.2$ and $q=0.3$. The PE-star partial densities 
    were chosen to be almost the same in both cases, i.e., 
    $\rho_{\rm s}\sigma_{\rm s}^3=0.27$ 
    and $\rho_{\rm s}\sigma_{\rm s}^3=0.24$, respectively. The corresponding colloid 
    densities are $\rho_{\rm c}\sigma_{\rm c}^3=0.29$ (main plot) and 
    $\rho_{\rm c}\sigma_{\rm c}^3=0.05$ (inset). Upon varying the size ratio, the peak 
    positions shift and the $(k=0)$-values of the partial structure factors shown change 
    significantly while there is no remarkable effect on the different peaks' height. 
    }
\end{center}
\end{figure}

Fig.\ \ref{fig:structure2} illustrates the typical shape of the partial structure factors 
$S_{ij}(k)$. Here, we chose the parameters as follows: 
the PE-star functionality is 
$f=18$, we set the size ratio to $q=0.2$, fixed the density of the stars as 
$\rho_{\rm s}\sigma_{\rm s}^3=0.12$, 
and considered several values of the colloidal 
density $\rho_{\rm c}\sigma_{\rm c}^3$. When comparing the three main plots of the 
figure, the first finding is that 
the locations of the different Lifshitz lines in 
density space strongly vary. These lines mark the 
respective structure factors' cross-over 
between a regime where they display a local minimum 
in the long wavelength limit $k\rightarrow 0$ 
and a region where the behaviour changes to developing 
a local maximum for the same $k$-values. 
While for the given amount of stars in the system the star--star Lifshitz line is obviously 
immediately crossed for practically arbitrary low colloid concentrations 
[fig.\ \ref{fig:structure2}(b)], we need an noticeably increased partial colloid density lying
in the range of about $\rho_{\rm c}\sigma_{\rm c}^3\approx 0.25\ldots 0.5$ for the colloid--star 
structure factor to experience such cross-over 
[fig.\ \ref{fig:structure2}(c)]. In case of the 
colloid--colloid structure factor, the corresponding values of the colloid density are 
even higher, about $\rho_{\rm c}\sigma_{\rm c}^3\approx 0.7$ for the parameters used here 
[fig.\ \ref{fig:structure2}(a)]. Another indication of the demixing transition we are searching 
for within the scope of this paper and that is expected to occur upon adding more and more 
colloids and stars to the binary mixture is the tendency of all partial structure factors to 
diverge in the aforementioned long wavelength limit, i.e., 
$S_{\rm cc}(k\rightarrow 0)\rightarrow+\infty$, $S_{\rm ss}(k\rightarrow 0)\rightarrow+\infty$ 
and $S_{\rm cs}(k\rightarrow 0)\rightarrow-\infty$, thus demonstrating that we approach 
the spinodal line. The inset in 
fig.\ \ref{fig:structure2}(b) was included in order to
again demonstrate the huge difference in the structural length scales of the two
species present in the mixture. The pre-peak in the
cross structure factor $S_{\rm cs}(k)$ is 
without any direct physical interpretation,
while pre-peaks in the intra-species structure 
factors would evince micro-phase separation
\cite{mossa:langmuir:04,sciortino:prl:04,sear:jcp:99,imperio:jcp:06,imperio:jpcm:04}. 
Since the latter peaks are completely absent in our case, we may once more conclude that the 
system is expected to macro-phase separate instead of forming clusters.

\begin{figure}[t]
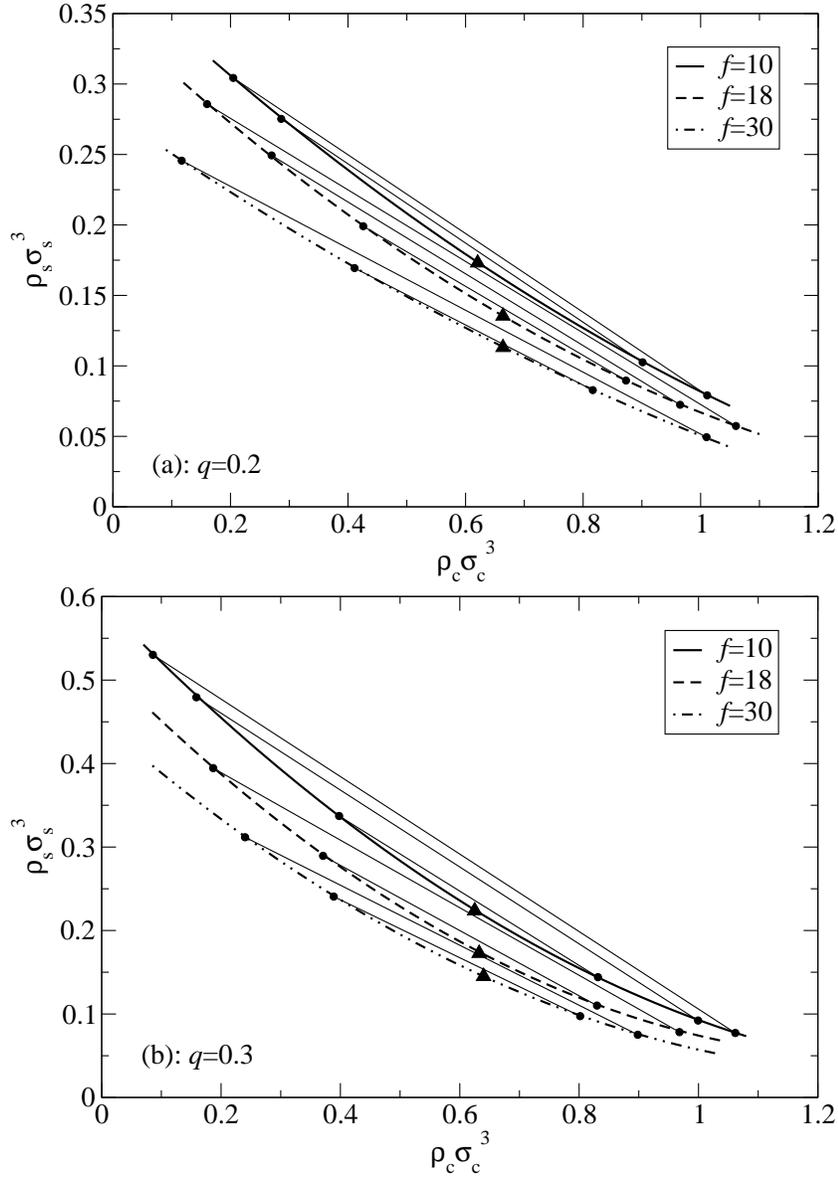

  \begin{center}
    \includegraphics[width=11cm,clip]{./figure9a.eps}
    \includegraphics[width=11cm,clip]{./figure9b.eps}
    \caption{\label{fig:binodal}
    Demixing binodals calculated according to the procedures introduced in the
    main text for (a) $q=0.2$ and (b) $q=0.3$, and different values
    of the PE-stars' functionality $f$. 
    In order to illustrate the coexisting colloid-poor
    and colloid-rich phases, we additionally show several tie lines. In this
    connection, please note that we in fact used much more such point pairs in order
    to obtain the binodal lines and not only the shown ones. Based on the full sets
    of coexisting fluid phases we computed, we made rough estimates for the positions of 
    the respective critical points in the $\rho_{\rm c}$-$\rho_{\rm c}$ plane, represented 
    by the filled triangles.
    }
\end{center}
\end{figure}

Finally, fig.\ \ref{fig:structure3} depicts 
the $q$-dependence of the cross structure
factors for $f=30$ and two different values of the colloid density
$\rho_{\rm c}\sigma_{\rm c}^3$ (main plot and inset). For both size ratios investigated, 
the star densities $\rho_{\rm s}\sigma_{\rm s}^3$ are chosen to be almost the same\footnote{
They are not exactly the same since such results are not systematically available due
to the fact that we originally solved the OZ relation together with the RY closure for 
points in the density plane where the star density takes 'smooth' values when scaled with 
respect to the colloidal diameter $\sigma_{\rm c}$, not their own diameter $\sigma_{\rm s}$.
}. As obvious from the plots, a change in $q$ only affects the peak positions and the depth
of the local minimum for $k\rightarrow 0$, but there is no significant influence on the peak 
heights of the functions. This is in agreement with the findings for the $q$-dependence 
of the effective colloid--colloid interactions, see fig.\ \ref{fig:veff}, and essentially means 
that the size ratio $q$ is crucial for determining the typical structural length scales,
but hardly for how pronounced this structure is. 

\subsection{Fluid-fluid phase equilibria}
\label{sec:phase}

After having found plenty of evidence in our hitherto analysis for a mixing--demixing 
transition taking place for certain ranges of partial densities $\rho_i\sigma_i^3$, we
finally come to a more quantitative description based on the corresponding binodals obtained
as explained above. In fig.\ \ref{fig:binodal} we show the obtained
demixing binodals 
for size ratios $q=0.2$ [fig.\ \ref{fig:binodal}(a)] and $q=0.3$ 
[fig.\ \ref{fig:binodal}(b)] 
and for different PE-star functionalities $f$, as denoted in the legend boxes.
In addition, we connected some of the point pairs used to compute the binodals 
and representing coexisting colloid-rich and colloid-poor phases by tie lines. 
Concerning
the mutual positions of the binodals in the density plane, it can be seen 
that they shift towards higher PE-star concentrations 
upon increasing the size ratio 
$q$ and/or decreasing the PE-stars' functionality $f$. 
This characteristic behaviour is in 
agreement with previous studies of binary mixtures of colloids and neutral polymer stars
\cite{dzubiella:jcp:116}. 
The filled triangles in fig.\ \ref{fig:binodal} denote rough 
estimates for the respective critical points' positions determined graphically by taking 
the tie lines into account. The critical points 
move towards slightly lower colloid 
densities when lowering the PE-stars' arm number, whereas there is no significant effect of
altering the size ratio. 

The star densities $\rho_{\rm s}\sigma_{\rm s}^3$ that bring 
about a demixing instability are
typically higher for the case $q=0.3$ than for the case $q=0.2$. This
looks counterintuitive at first sight, since one expects that larger
PE-stars will destabilise the mixture earlier. In order to put the
numbers in their appropriate context, it is useful to employ the 
picture of the effective colloid-colloid potential, which includes
a star-induced attraction. Here, the range and depth of this attraction
steer the occurrence of the demixing binodal, which is equivalent to
a separation between a colloidal fluid and a colloidal gas. The natural
length scale in this picture is the colloid diameter $\sigma_{\rm c}$;
concomitantly, the physically relevant density in making comparisons
between the $q=0.2$ and the $q=0.3$ cases should be  
scaled with the colloid size: 
$\rho_{\rm s}\sigma_{\rm c}^3 = q^{-3}\rho_{\rm s}\sigma_{\rm s}^3$.
It can be easily seen that the additional prefactor $q^{-3}$ renders
the rescaled star densities for $q=0.3$ indeed lower than the ones
for $q=0.2$, in agreement with the intuitive expectations.

The volume terms for the integrated out 
counterions \cite{hansen:97,hansen:99,denton:00}
do not affect the phase boundaries, since,
under the assumption of full absorbing in the stars'
interiors, they are simply proportional
to the number $N_{\rm s}$ of the latter \cite{hoffmann:jcp:04}
and thus they cause a trivial
shift of the stars' chemical potential, without affecting the
solution's osmotic pressure \cite{likos:pr:01}.
Finally, we mention that we did not consider the
competition between the demixing binodals and the crystallization of
the colloids. 
The investigation of the system's solid states lies beyond
the scope of this work. The trends 
found for the $f$- and $q$-dependences are
comparable to the colloid--star polymer case 
mentioned above. Although the underlying pair
potentials are different to a certain degree, 
a closer inspection to the full phase
diagrams in Ref.\ \cite{dzubiella:jcp:116} can give hints
regarding the stability of the binodals against preemption
by the freezing lines. Provided the positions
of the freezing lines are not too different here, it seems to be reasonable to 
assume based on such a comparison that our demixing 
lines will survive at least for the larger size ratio between stars
and colloids. Nevertheless, the existence of a demixing binodal,
even in the case that the latter is preempted by crystallization,
has important consequences for the time scales involved in the
dynamics of crystallization \cite{frenkel1,frenkel2}.

\section{Summary and conclusions}
\label{sec:summary}

We have put forward a coarse-grained description of mixtures
between neutral, spherical, hard colloids and multiarm polyelectrolyte
stars of size smaller than the colloidal particles. Effective 
interactions between the constituent particles have been
employed throughout, allowing for a mesoscopic description that
leads to valuable information on the structure and thermodynamics
of the two-component mixture. The cross interaction, which has
been derived in this work, is sufficiently repulsive to bring
about regions of instability in the phase diagram and leading
thereby to macroscopic, demixing phase behaviour. This, in turn,
can be rationalised by means of the depletion potentials between
the colloids, which are induced by the stars, and feature
attractive tails akin to those encountered in usual colloid-polymer
mixtures.

The form of the cross interaction plays a crucial role in determining
stability and can, by suitable tuning, completely change
the behaviour of the mixture from macroscopic phase separation
to microphase structuring with a finite wavelength. In this
respect, a very promising direction of investigation is to
allow for the colloids to carry a charge {\it opposite} to that
of the arms of the polyelectrolyte stars. Preliminary results
already indicate a rich variety of resulting complexation
morphologies between the two constituents \cite{kon:unpub}. 
A detailed investigation of the complexation characteristics and the
morphologies of the ensuing macroscopic phases is the
subject of ongoing work.

\section*{Acknowledgments}
The authors wish to thank Joachim Dzubiella 
and Christian Mayer for helpful discussions.



\section*{References}
\begin{thebibliography}{99}

\bibitem{pincus:macromol:24}
Pincus P 1991 {\it Macromolecules} {\bf 24} 2912

\bibitem{wang:pre:04} Wang H and Denton A R 2004 {\it Phys.\ Rev.\ E} {\bf 70}
041404

\bibitem{borisov:97} Borisov O V and Zhulina E B 1997 {\it J.\ Phys.\ II (Paris)}
{\bf 7} 499

\bibitem{borisov:98} Borisov O V and Zhulina E B 1998 {\it Eur.\ Phys.\ J.\ B}
{\bf 4} 205

\bibitem{klein:99} Klein Wolterink J, Leermakers F A M, Fleer G J,
Koopal L K, Zhulina E B and Borisov O V 1999 {\it Macromolecules} {\bf 32}
2365

\bibitem{klein:02} Klein Wolterink J, van Male J, Cohen Stuart M A,
Koopal L K, Zhulina E B and Borisov O V 2002 {\it Macromolecules} {\bf 35}
9176

\bibitem{jusufi:prl:02}
Jusufi A, Likos C N and L\"owen H 2002
{\it Phys.\ Rev.\ Lett.} {\bf 88} 018301

\bibitem{jusufi:jcp:02}
Jusufi A, Likos C N and L{\"o}wen H 2002
{\it J.\ Chem.\ Phys.} {\bf 116} 11011 

\bibitem{denton:03} Denton A R 2003 {\it Phys.\ Rev.\ E} {\bf 67} 011804

\bibitem{hoffmann:jpcm:03}
Likos C N, Hoffmann N, Jusufi A and L\"owen H 2003,
{\it J.\ Phys.: Condens.\ Matter} {\bf 15} S233 

\bibitem{hoffmann:jcp:04}
Hoffmann N, Likos C N and L\"owen H 2004
{\it J.\ Chem.\ Phys.} {\bf 121} 7009 

\bibitem{ishizu:05} Furukawa T and Ishizu K 2005 {\it Macromolecules} {\bf 38}
2911

\bibitem{stamm:03} Gorodyska G, Kiriy A, Minko S, Tsitsilianis C and
Stamm M 2003 {\it Nano Letters} {\bf 3} 365

\bibitem{serpe:04} Serpe M J, Kim J and Lyon L A 2004 {\it Adv.\ Mater.}
{\bf 16} 184

\bibitem{kim:05} Kim J, Serpe M J and Lyon L A 2005 
{\it Angew.\ Chem.\ Int.\ Ed.} {\bf 44} 1333

\bibitem{konieczny:jcp:06}
Konieczny M and Likos C N 2006
{\it J.\ Chem.\ Phys.} {\bf 124} 214904

\bibitem{tuinier:jpcm:05} Fortini A, Dijkstra M and Tuinier R 2005
{\it J. Phys.: Condens.\ Matter} {\bf 17} 7783

\bibitem{jusufi:cps:04} Jusufi A, Likos C N and Ballauff M 2004
{\it Colloid Polym.\ Sci.} {\bf 282} 910

\bibitem{manning:jcp:69} Manning G S 1969 {\it J.\ Chem.\ Phys.} {\bf 51} 
924

\bibitem{konieczny:jcp:121} 
Konieczny M, Likos C N and L\"owen H 2004
{\it J.\ Chem.\ Phys.} {\bf 121} 4913 

\bibitem{konieczny:molsym}
Konieczny M and Likos C N 2006 
{\it Macromolecular Symposia} in press

\bibitem{jusufi:jpcm:13} 
Jusufi A, Dzubiella J, Likos C N, von
Ferber C and L\"owen H 2001
{\it J.\ Phys.: Condens.\ Matter} {\bf 13} 6177

\bibitem{hansen:book} Hansen J-P and McDonald I R
{\it Theory of Simple Liquids}, 2nd ed. (London: Academic Press 1986)

\bibitem{hansen:arpc:51} Hansen  J-P and L\"owen H 2000 
{\it Annu.\ Rev.\ Phys.\ Chem.} 
{\bf 51} 209 

\bibitem{lebowitz:jcp}
Lebowitz J L and Rowlinson J S 1964 
{\it J.\ Chem.\ Phys.} {\bf 41} 133

\bibitem{rogers:pra:84}
Rogers  F J and Young D A 1984 {\it Phys.\ Rev.\ A} {\bf 30} 999

\bibitem{biben:jpcm:91}
Biben T and Hansen J P 1991
{\it J.\ Phys.: Condens.\ Matter} {\bf 3} 65

\bibitem{kirkwood:jcp:51} Kirkwood J G and Buff F P 1951 
{\it J.\ Chem.\ Phys.} {\bf 19} 774

\bibitem{ashcroft:78} Ashcroft N W and Stroud D 1978 
{\it Solid State Phys.} {\bf 33} 2 (Academic: New York)

\bibitem{likos:jcp:92}
Likos C N and Ashcroft N W 1992 {\it J.\ Chem.\ Phys.} {\bf 97} 9303

\bibitem{biben:prl:91}
Biben T and Hansen J-P 1991 {\it Phys.\ Rev.\ Lett.} {\bf 66} 2215

\bibitem{bhatia:pre:70}
Bathia A B and Thornton D E 1970 {\it Phys.\ Rev.\ B} {\bf 2} 3004 

\bibitem{dzubiella:jcp:116}
Dzubiella J, Likos C N and L\"owen H 2002
{\it J.\ Chem.\ Phys.} {\bf 116} 9518

\bibitem{row:book} Rowlinson J S and Swinton F L 
{\it Liquids and Liquid Mixtures} (London: Butterworth 1982)

\bibitem{landau:book}
Landau  L D and Lifshitz E M {\it Course of Theoretical Physics:
Statistical Physics}, 3rd ed. (Oxford: Pergamon Press 1980)

\bibitem{carnahan:jcp:69}
Carnahan N F and Starling K E 1969 {\it J.\ Chem.\ Phys.} {\rm 51}
635

\bibitem{andy:02} Archer A J, Likos C N and Evans R 2002 
{\it J.\ Phys.: Condens.\ Matter} {\bf 14} 12031

\bibitem{mayer:pre:70}
Mayer C, Likos C N and L\"owen H 2004
{\it Phys.\ Rev.\ E} {\bf 67} 0414025

\bibitem{dzubiella:epl}
Dzubiella J, Likos C N and L\"owen H 2002
{\it Europhys.\ Lett.} {\bf 58} 133

\bibitem{mendez:pre:00}
M{\'e}ndez-Alcaraz J M and Klein R 2000 {\it Phys.\ Rev.\ E} {\bf 61}
4095

\bibitem{koenig:pre:01}
K\"onig A and Ashcroft N W 2001 {\it Phys.\ Rev.\ E} {\bf 63}
041203 

\bibitem{damico:physa:97}
D'Amico I and L\"owen H 1997 {\it Physica A} {\bf 237} 25 

\bibitem{allahyarov:prl:98}
Allahyarov E, D'Amico I and L\"owen H 1998 {\it Phys.\ Rev.\ Lett.} {\bf 81}
{\bf 81} 1334

\bibitem{allahyarov:jpcm:01}
Allahyarov E and L\"owen H 2001 {\it J.\ Phys.: Condens.\ Matter} {\bf 13}
L277

\bibitem{attard:jcp:89}
Attard P 1989 {\it J.\ Chem.\ Phys.} {\bf 91} 3083

\bibitem{moreno:pre:06}
Moreno A J and Colmenero J 2006 {\it Phys.\ Rev.\ E} {\bf 74}
021409

\bibitem{mossa:langmuir:04}
Mossa S, Sciortino F, Tartaglia P and Zaccarelli E 2004
{\it Langmuir} {\bf 20} 10756

\bibitem{sciortino:prl:04}
Sciortino F, Mossa S, Zaccarelli E and Tartaglia P 2004
{\it Phys.\ Rev.\ Lett.} {\bf 93} 055701

\bibitem{sear:jcp:99}
Sear R P and Gelbart W M 1999 {\it J.\ Chem.\ Phys} {\bf 110}
4582

\bibitem{imperio:jcp:06}
Imperio A and Reatto L 2006 {\it J.\ Chem.\ Phys.} {\bf 124}
164712 

\bibitem{imperio:jpcm:04}
Imperio A and Reatto L 2004 {\it J.\ Phys.: Condens.\ Matter} {\bf 16}
3769 

\bibitem{vink:pre:05}
Vink R L C, Jusufi A, Dzubiella J, and Likos C N 2005 {\it Phys.\ Rev.\ E}
{\bf 72} 030401(R)

\bibitem{loverso:jpcm:05}
Lo Verso F, Pini D and Reatto L 2005
{\it J.\ Phys.: Condens.\ Matter} {\bf 17} 771

\bibitem{loverso:pre:06}
Lo Verso F, Vink R L C, Pini D and Reatto L 2006
{\it Phys.\ Rev.\ E} {\bf 73} 061407

\bibitem{hansen:97} van Roji R and Hansen J-P 1997 {\it Phys.\ Rev.\ Lett.}
{\bf 79} 3082

\bibitem{hansen:99} van Roji R, Dijkstra M and Hansen J-P 1999
{\it Phys.\ Rev.\ E} {\bf 59} 2010

\bibitem{denton:00} Denton A R 2000 {\it Phys.\ Rev.\ E} {\bf 62} 3855

\bibitem{likos:pr:01} Likos C N 2001 {\it Phys. Rep.} {\bf 348} 267

\bibitem{frenkel1} ten Wolde P R and Frenkel D 1997 {\it Science} {\bf 277}
1975

\bibitem{frenkel2} Frenkel D 1999 {\it Physica A} {\bf 263} 26

\bibitem{kon:unpub} Konieczny M, Jusufi A and Likos C N unpublished

\end{thebibliography}
\end{document}